\documentclass[a4paper,fleqn,usenatbib,useAMS]{mnras}

\usepackage{graphicx}	
\usepackage{amsmath}	
\usepackage{amssymb}	
\usepackage{multicol}        
\usepackage{bm}		
\usepackage{pdflscape}	

\newcommand{\Deltas}{\Delta^2_{\mathrm{s}}(\mathbf{k})}
\newcommand{\Deltar}{\Delta^2_{\mathrm{r}}(\mathbf{k})}
\newcommand{\Deltasav}{\Delta^2_{\mathrm{s}}(k)}
\newcommand{\Deltarav}{\Delta^2_{\mathrm{r}}(k)}
\newcommand{\Deltawindow}{\Delta^2_{\mathrm{s,window}}(k)}

\usepackage[T1]{fontenc}
\usepackage{ae,aecompl}

\usepackage{times}

\title[The wedge bias]{The wedge bias in reionization 21-cm power spectrum measurements}

\author[H. Jensen et al]{Hannes Jensen$^{1,2}$\thanks{Contact e-mail: \href{hjens@astro.su.se}{hjens@astro.su.se}}, Suman Majumdar$^{1}$, Garrelt Mellema$^{1}$, Adam Lidz$^{3}$, \newauthor
Ilian T.\ Iliev$^{4}$, Keri L.\ Dixon$^{4}$
\\
$^{1}$Department of Astronomy and Oskar Klein Centre, Stockholm University, AlbaNova, SE-10691 Stockholm, Sweden \\
$^{2}$Department of Physics and Astronomy, Uppsala University, Uppsala, Sweden \\
$^{3}$Department of Physics and Astronomy, University of Pennsylvania, Philadelphia, PA 19104, USA \\
$^{4}$Astronomy Centre, Department of Physics \& Astronomy, Pevensey II Building, University of Sussex, Falmer, Brighton BN1 9QH, UK} 

\date{\today}

\pubyear{2015}

\begin{document}
\label{firstpage}
\pagerange{\pageref{firstpage}--\pageref{lastpage}}
\maketitle
\begin{abstract}
A proposed method for dealing with foreground emission in upcoming 21-cm observations from the epoch of reionization is to limit observations to an uncontaminated window in Fourier space. Foreground emission can be avoided in this way, since it is limited to a wedge-shaped region in $k_{\parallel}, k_{\perp}$ space. However, the power spectrum is anisotropic owing to redshift-space distortions from peculiar velocities. Consequently, the 21-cm power spectrum measured in the foreground avoidance window---which samples only a limited range of angles close to the line-of-sight direction---differs from the full redshift-space spherically-averaged power spectrum which requires an average over \emph{all} angles. In this paper, we calculate the magnitude of this ``wedge bias'' for the first time. We find that the bias amplifies the difference between the real-space and redshift-space power spectra. The bias is strongest at high redshifts, where measurements using foreground avoidance will over-estimate the redshift-space power spectrum by around 100 per cent, possibly obscuring the distinctive rise and fall signature that is anticipated for the spherically-averaged 21-cm power spectrum. In the later stages of reionization, the bias becomes negative, and smaller in magnitude ($\lesssim 20$ per cent). 
\end{abstract}




\section{Introduction}

A major obstacle facing any upcoming measurement of the 21-cm signal from the epoch of reionization (EoR) is foreground emission from galactic and extra-galactic sources (e.g.\ \citealt{jelic2010}). Two separate approaches for dealing with this problem have been proposed: foreground removal and foreground avoidance. Foreground removal involves modeling the foreground emission in order to subtract it from the total signal (e.g.\ \citealt{chapman2013,chapman2014}). In foreground avoidance, the idea is instead to use only the parts of the signal in $k$ space (Fourier space) that are unaffected by foregrounds.

It has been shown that foreground contamination of the EoR 21-cm signal will be localized to a wedge-shaped region in cylindrical $k_{\parallel},k_{\perp}$ space, at low values of $\mu\equiv k_{\parallel} /|\mathbf{k}|$ \citep{datta2010,trott2012,dillon2014,pober2014}. This leaves an ``EoR window'' at high $\mu$, where the pristine signal can be observed. Limiting observations to this window has the advantage of not requiring any knowledge of the properties of the foreground emission. The main downside is that a (potentially large) part of the signal has to be thrown away, resulting in lower signal-to-noise. A second downside is that direction dependent quantities, such as the $\mu$-dependent power spectrum become hard to measure \citep{pober2015}. 

Even for measurements of the spherically-averaged power spectrum (which is not dependent on $\mu$), one must be careful when comparing observations to theory. Observations can only be done in redshift space, where the apparent positions of emitters are distorted by their peculiar velocities along the line-of-sight, and it is well-known that the redshift-space power spectrum, $\Deltas$, can differ substantially from the real-space power spectrum, $\Deltar$, even after spherical averaging \citep{mao12}\footnote{We show our results in terms of the ``dimensionless'' 21-cm power spectrum $\Delta^2(k) \equiv k^3 P(k)/(2\pi^2)$. We use subscript $r$ to denote real-space quantities, and $s$ to denote redshift-space quantities.}. However, since the redshift-space power spectrum is not spherically symmetric, limiting a measurement to an EoR window---i.e.\ including only some values of $\mu$---will introduce an additional bias. We then have to distinguish between three different power spectra: $\Deltarav$, which is the fundamental one, and the raw output from most simulations; $\Deltasav$, which is the quantity seen by an observer; and $\Deltawindow$, which is a biased estimator of $\Deltasav$.

The bias that makes $\Deltawindow$ different from $\Deltas$ has been hinted at in the literature (e.g. \citealt{liu2014}), but to our knowledge, it has never been quantified. Here, we calculate the magnitude of the bias and discuss how to correct for it when interpreting observations and simulations.
%
%

\section{Theory}
\subsection{The wedge}
The foregrounds that contaminate EoR 21-cm measurements are expected to be spectrally smooth, which means they should only affect the lowest $k_{\parallel}$ modes. However, because of the frequency-dependence of an interferometer's response, the contamination will leak into higher $k$ modes. This effect, known as ``mode mixing'', is strongest for long baselines (high values of $k_{\perp}$) which have higher fringe rates. Mode mixing causes the foregrounds to become confined to a wedge-shaped region in $k_{\parallel},k_{\perp}$ space \citep{datta2010,trott2012,pober2014,dillon2014,liu2014,pober2015}.

The contamination by flat-spectrum foreground sources can be shown to extend no farther than the line defined by the following relation:
\begin{equation}
k_{\parallel, \mathrm{min}} = \left( \sin{\theta_{\mathrm{FoV}}}\frac{D_\mathrm{M}(z)E(z)}{D_\mathrm{H}(1+z)} \right) k_{\perp} \equiv C k_{\perp},
\label{eq:wedge_slope}
\end{equation}
where $\theta_{\mathrm{FoV}}$ is the angular radius of the field-of-view, $D_{\mathrm{H}} \equiv c/H_0$, $E(z) \equiv \sqrt{\Omega_{\mathrm{m}}(1+z)^3 + \Omega_{\mathrm{\Lambda}}}$ and $D_{\mathrm{M}}(z)$ is the transverse comoving distance, which for a flat cosmology is equivalent to the comoving distance: $D_{\mathrm{M}}(z) = D_{\mathrm{H}} \int_0^z \mathrm{d}z'/E(z')$, (e.g.\ \citealt{parsons2012,thyagarajan2013,dillon2014}).
%
%
%
%

From the slope of the wedge defined in Equation \eqref{eq:wedge_slope} we get the minimum value of $\mu$ that can be observed without foreground contamination:
%
%
\begin{equation}
\mu_{\mathrm{min}} = \frac{C}{\sqrt{C^2+1}}.
\end{equation}
Unfortunately, there is some lack of consensus as to what value for $\theta_{\mathrm{FoV}}$  should be used when calculating the wedge slope. The most pessimistic assumption would be to include contamination from sources on the horizon, and set $\theta_{\mathrm{FoV}}=90$ degrees, although observations seem to indicate that some ``supra-horizon'' contamination may occur beyond this limit, possibly since real foregrounds are in fact not spectrally flat \citep{pober13}. On the other hand, it has been argued that it is possible to avoid contamination from sources outside the primary beam of the telescope, which would make the EoR window significantly larger \citep{pober2014}. 
%
%
%
%

\begin{figure}
\centering
\includegraphics[width=0.4\textwidth]{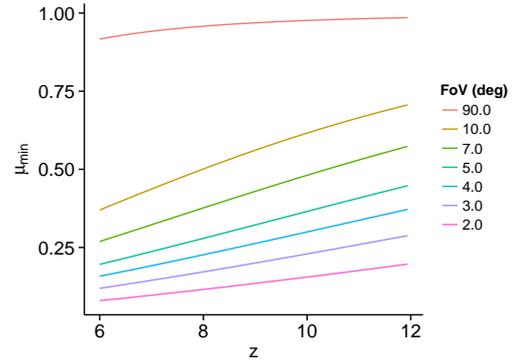}
\caption{The smallest $\mu$ that is observable using foreground avoidance, as a function of redshift for different field-of-view sizes. A 90 degree field-of-view corresponds to the horizon limit.}
\label{fig:mu_hor}
\end{figure}

In Figure \ref{fig:mu_hor}, we show $\mu_{\mathrm{min}}$ as a function of redshift for a few different field-of-view sizes. Our aim with this paper is not to provide a realistic model for the EoR window, so we will simply show our results for the optimistic case of $\mu_{\mathrm{min}}=0.5$ and the more pessimistic case of $\mu_{\mathrm{min}}=0.95$.

\begin{figure}
\includegraphics[width=0.5\textwidth]{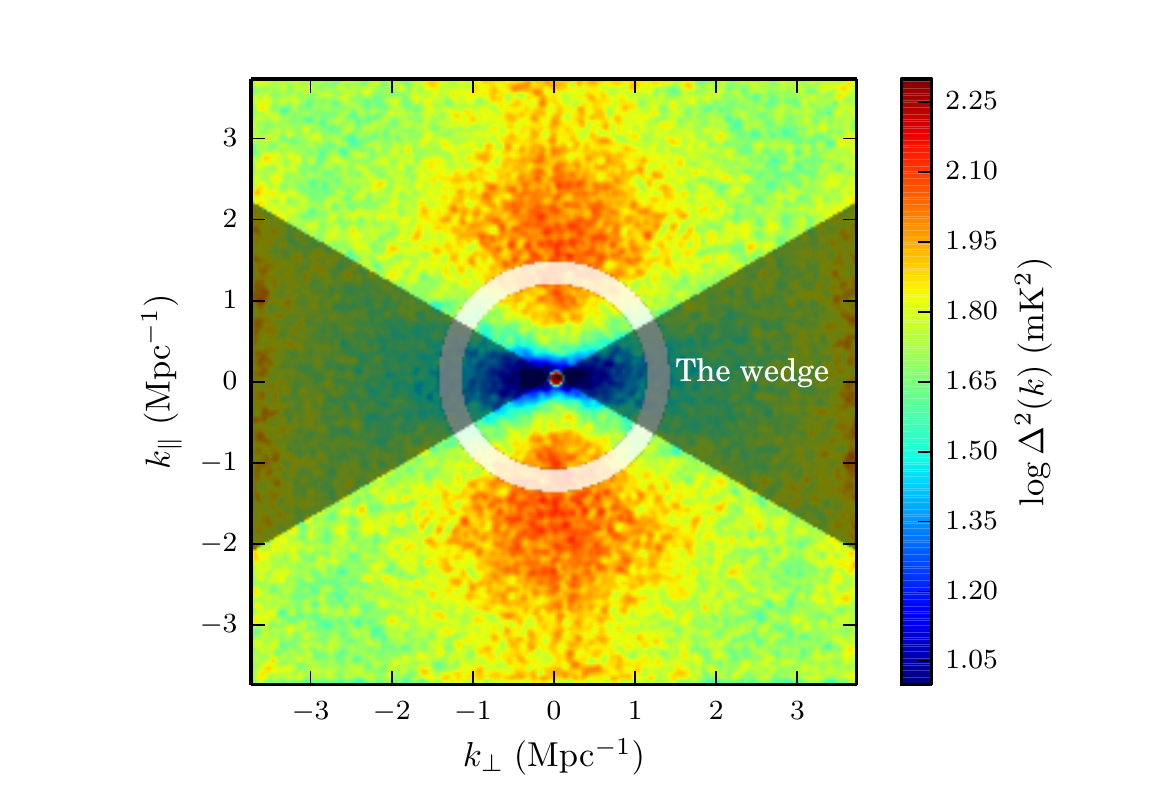}
\caption{A slice from the full three-dimensional 21-cm power spectrum at $z=10$, smoothed with a Gaussian kernel for visual clarity. The shaded black area shows the region where $\mu<0.5$. The  white region shows an annulus of constant $k$. Notice that the power spectrum is not constant within this annulus.}
\label{fig:ps_slice}
\end{figure}
In addition to the wedge, the EoR window is also restricted by foregrounds on all low $k_{\parallel}$ modes. This edge is commonly assumed to be around $k_{\parallel} = 0.05$ Mpc$^{-1}$ \citep{dillon2014}. The lowest $k$ mode we consider in this paper is $k=0.09$ Mpc$^{-1}$ which is well within the window. First-generation telescopes such as LOFAR and MWA are expected to be most sensitive to the EoR 21-cm power spectrum on scales around $k\approx 0.1$ Mpc$^{-1}$ \citep{harker2010,beardsley2013}.

\subsection{Redshift space and the wedge bias}
The real-space 21-cm signal---which is the typical output from reionization simulations---has no direction, and its power spectrum is isotropic. However, the real-space signal can never be observed. Any observation will see the signal in redshift space, where the apparent positions of emitters are distorted along the line-of-sight by their peculiar velocities (see e.g.\ \cite{mao12} for an overview of the effects of redshift-space distortions on the EoR 21-cm signal). An emitter with a line-of-sight peculiar velocity $v_{\parallel}$ and real-space position $\mathbf{r}$ will be translated to an apparent, redshift-space position $\mathbf{s}$:
\begin{equation}
	\label{eq:redshift_translation}
	{\mathbf s} = {\mathbf r} + \frac{1 + z}{H(z)}v_{\parallel} \hat{r}.
\end{equation}

This distortion has two effects. First, it changes the contrast of the signal, making $\Deltasav$ differ from $\Deltarav$. Second, since the translation happens along the line-of-sight, it makes the redshift-space signal anisotropic. The same holds for the redshift-space power spectrum, $\Deltas$; it is now a function of $\mu$. In other words, to measure the spherically-averaged redshift-space power spectrum $\Deltasav$, we need to average over an entire spherical shell in $k$ space, including all values of $\mu$. Averaging only in the parts of a shell that are located in the EoR window will result in a biased measurement of $\Deltasav$. 

Figure \ref{fig:ps_slice}, shows a 2D slice from a simulated 3D power spectrum in redshift space. This figure makes apparent that the power spectrum is not isotropic and that averaging only outside the wedge will not accurately estimate $\Deltasav$. In the following sections we will investigate the magnitude of this error, which we will refer to as \emph{wedge bias}, and define as $[\Deltawindow-\Deltasav]/\Deltasav$, where $\Deltawindow$ is the power spectrum measured in the EoR window defined by some $\mu_{\mathrm{min}}$.

\section{Simulations}
\label{sec:simulations}
To study the wedge bias on the 21-cm power spectrum, we use a set of semi-numerical reionization simulations carried out on top of a numerical $N$-body simulation. The $N$-body simulation was performed with \textsc{cubep$^3$m} \citep{harnoisderaps13}. This code calculates gravitational forces on a particle-particle basis for small distances and on a mesh for longer distances. We used 6912$^3$ particles of mass $4\times 10^7 \; M_{\odot}$ on a 13824$^3$ mesh, which was later downsampled to 600$^3$ cells. The total size of the simulation volume was $500/h=714$ cMpc along each side. The minimum dark matter halo mass used in the simulation was $2.02 \times 10^9\;M_{\odot}$. 
%
%
%
%

The reionization part of the simulations was carried out with a modified version of the semi-numerical code described in \cite{choudhury09b} and \cite{majumdar14}. This code calculates the ionization state of the intergalactic medium by comparing the average number of ionizing photons entering in a cell with the average number of neutral hydrogen atoms in it. A cell is considered ionized if it is possible to find a sphere of some radius (limited by the assumed mean free path of the ionizing photons) around it within which the number of ionizing photons exceeds the number of neutral hydrogen atoms.

We use two different models for assigning ionizing fluxes to our dark matter halos. In the fiducial model, the ionizing flux is proportional to the halos mass, $M_{\mathrm{h}}$:
\begin{equation}
N_{\gamma}(M_{{\rm h}}) = N_{{\rm ion}} \frac{M_{{\rm h}} \Omega_{{\rm b}}}{m_{{\rm p}} \Omega_{{\rm m}}},
\label{eq:ionizing_flux}
\end{equation}
where $N_{\mathrm{ion}}$ is the number of photons entering the IGM per baryon in collapsed objects and $m_{{\rm p}}$ is the mass of a hydrogen atom. The total number of ionizing photons is not conserved in this scheme due to the overlapping of ionized regions \citep{zahn07}. We tune the value of $N_{\mathrm{ion}}$ at different redshifts to compensate for this and also to make sure that the resulting reionization history follows the same trend as the evolution of the mass averaged collapsed fraction with redshift. Our results (Section \ref{sec:results}) are sensitive mainly to neutral fraction rather than redshift, and so we will present our results as a function of $\bar{x}_{\mathrm{HI}}$ for the remainder of the paper.

\begin{figure}
\centering
\includegraphics[width=0.3\textwidth]{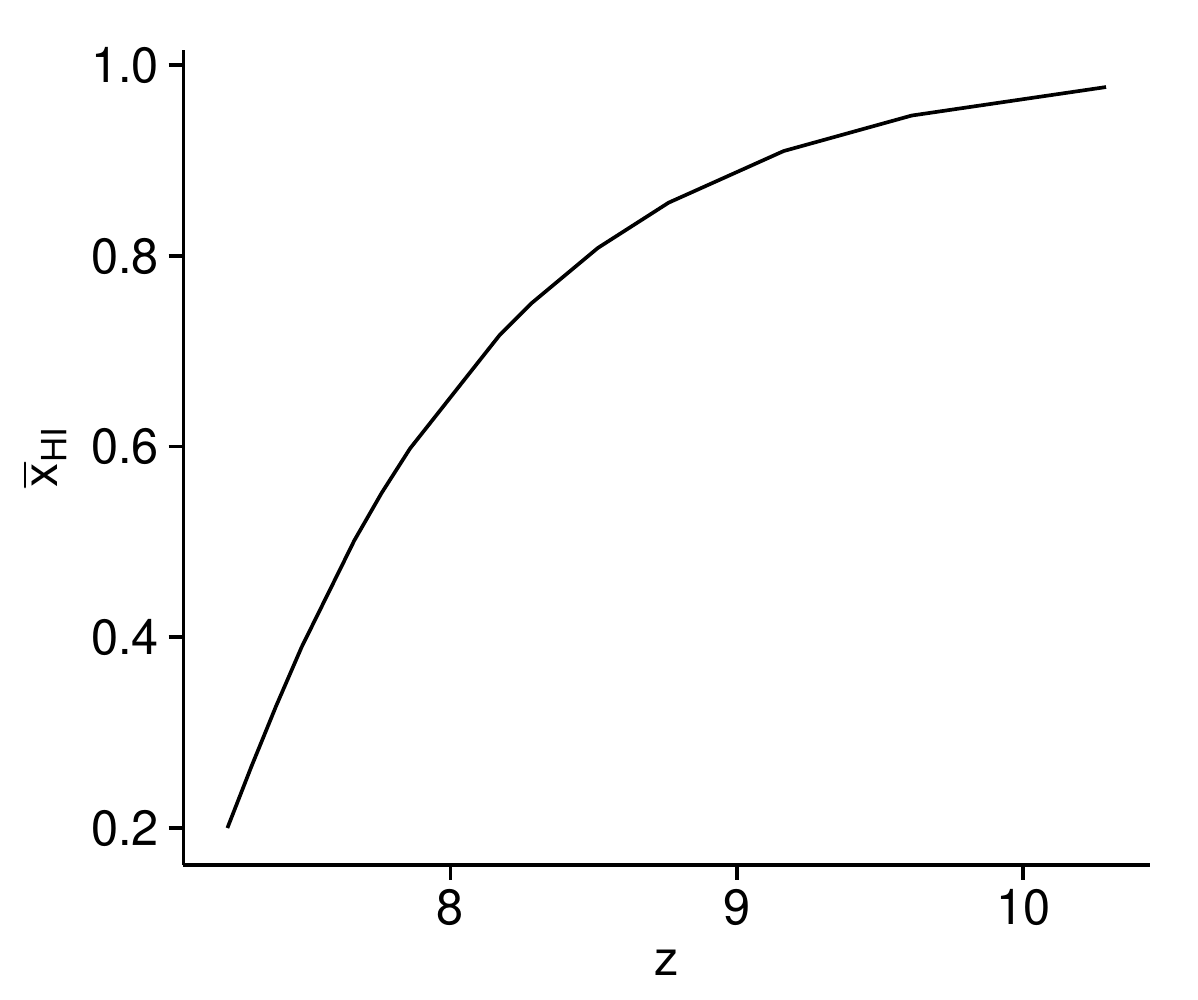}
\caption{The mass averaged  mean neutral fraction as a function of redshift in our simulations.}
\label{fig:xi_history}
\end{figure}

The second model, which we will refer to as the ``massive sources'' model, has $N_{\gamma} \propto M_{\mathrm{h}}^2$, with the proportionality constant tuned to give the same reionization history as the fiducial model. This model assigns higher fluxes to more massive sources, resulting in fewer and larger ionized bubbles. These two models were chosen to provide two very different examples of reionization topologies. The resulting reionization history (which is the same for both models) is shown in Figure \ref{fig:xi_history}.

After running the reionization simulations, we combine the density fields from the $N$-body simulations with the ionization fields to get the 21-cm brightness temperature, making the simplifying assumption that the spin temperature is much higher than the temperature of the cosmic microwave background (CMB). Finally, we convert the output from real space to redshift space, using the methodology described in \cite{jensen13}. All simulations use the cosmological
parameters from WMAP five year data release: $h = 0.7$, $\Omega_{\mathrm{m}} =
0.27$, $\Omega_{\Lambda} = 0.73$, $\Omega_{\mathrm{b}} h^2 = 0.0226$ \citep{komatsu09}.
%
%
%
%

\section{Results}
\label{sec:results}
To illustrate the bias that occurs when measuring the spherically-averaged redshift-space 21-cm power spectrum in the EoR window, we calculate the power spectrum from our simulated 21-cm volumes, both for the full volume and for a window defined by some value of $\mu_{\mathrm{min}}$. 

\begin{figure}
\includegraphics[width=0.5\textwidth]{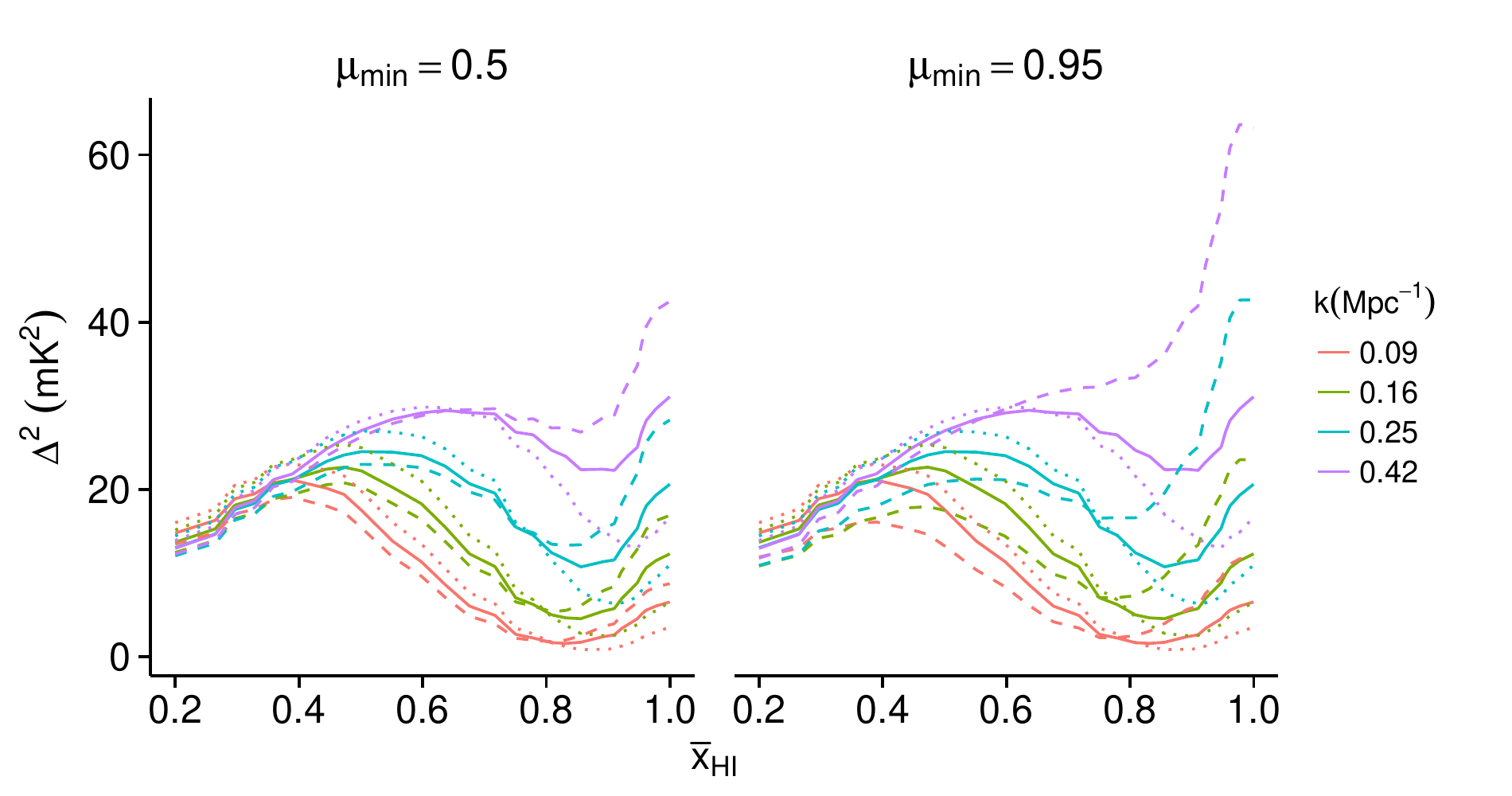}
\caption{The difference between $\Deltasav$, $\Deltarav$ and $\Deltawindow$. Dotted lines show the real-space power spectrum, $\Deltarav$, while solid lines show the redshift-space power spectrum $\Deltasav$. Dashed lines show $\Deltawindow$ for $\mu_{\mathrm{min}}=0.5$ (left panel) and $\mu_{\mathrm{min}}=0.95$ (right panel). All curves are for the fiducial model as a function of global neutral fraction.}
\label{fig:ps_fidu}
\end{figure}
%
%
%
%
Figure \ref{fig:ps_fidu} shows the power spectra for the fiducial model, in real space and redshift space both for an EoR window and for a full range of $\mu$ values. It is clear from this figure that when we restrict the measurement to certain $\mu$ values, the wedge bias increases the difference between $\Deltasav$ and $\Deltarav$. Since the redshift-space distortion effect varies with redshift (see e.g.\ \citealt{mao12}), so will the wedge bias. The bias is most pronounced on large scales (small values of $k$) and, as expected, becomes more significant in the case of a smaller EoR window (higher $\mu_{\mathrm{min}}$).  Problematically, increased bias at high redshift obscures the characteristic rise and fall signature in the 21-cm power spectrum.
%
%

\begin{figure}
\includegraphics[width=0.5\textwidth]{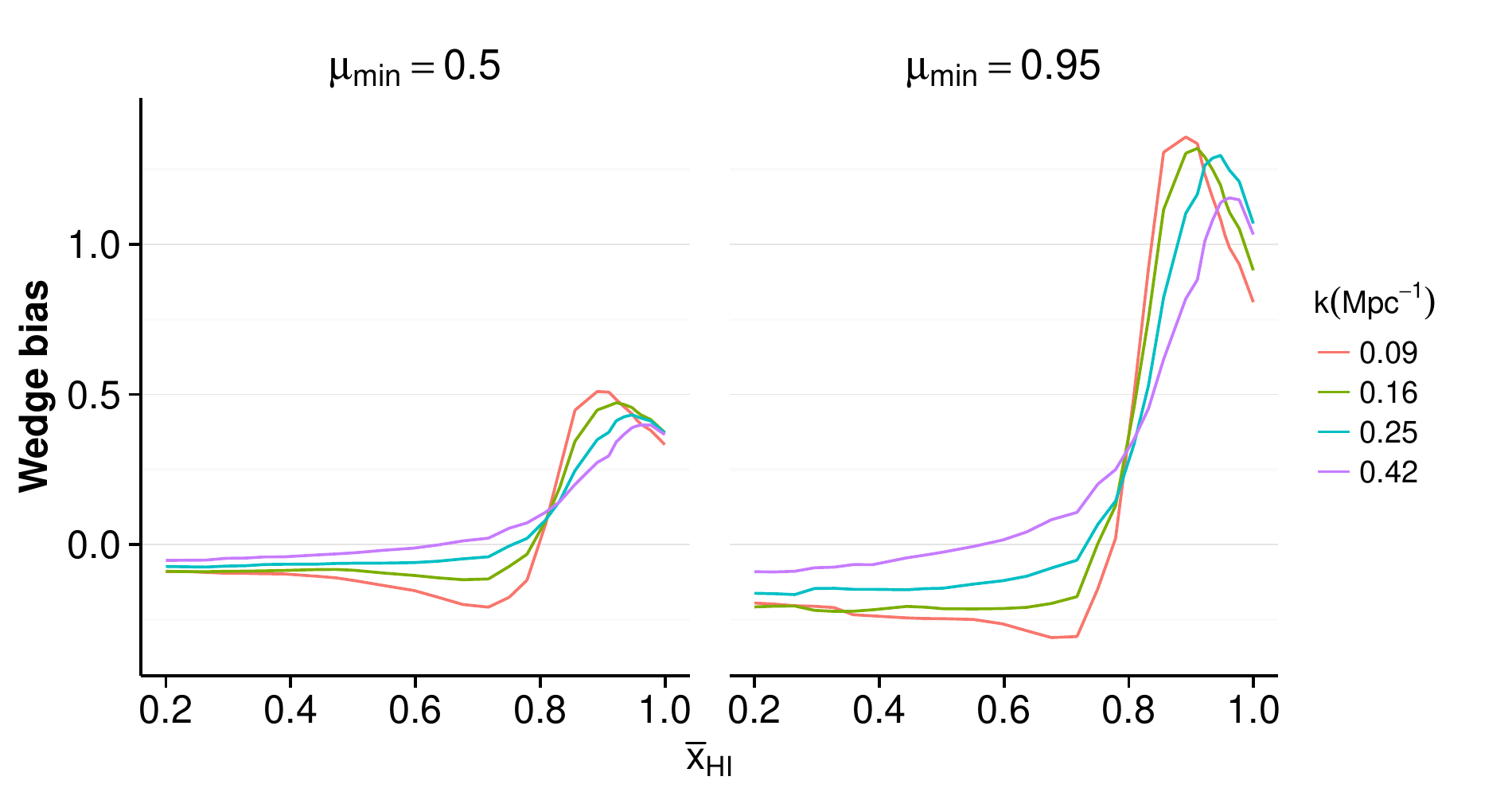}
\caption{The bias when measuring the spherically-averaged redshift-space power spectrum inside the EoR window, for a number of different $k$ modes and two values of $\mu_{\mathrm{min}}$. The results shown are for the fiducial model.}
\label{fig:frac_error_fidu}
\end{figure}

\begin{figure}
\includegraphics[width=0.5\textwidth]{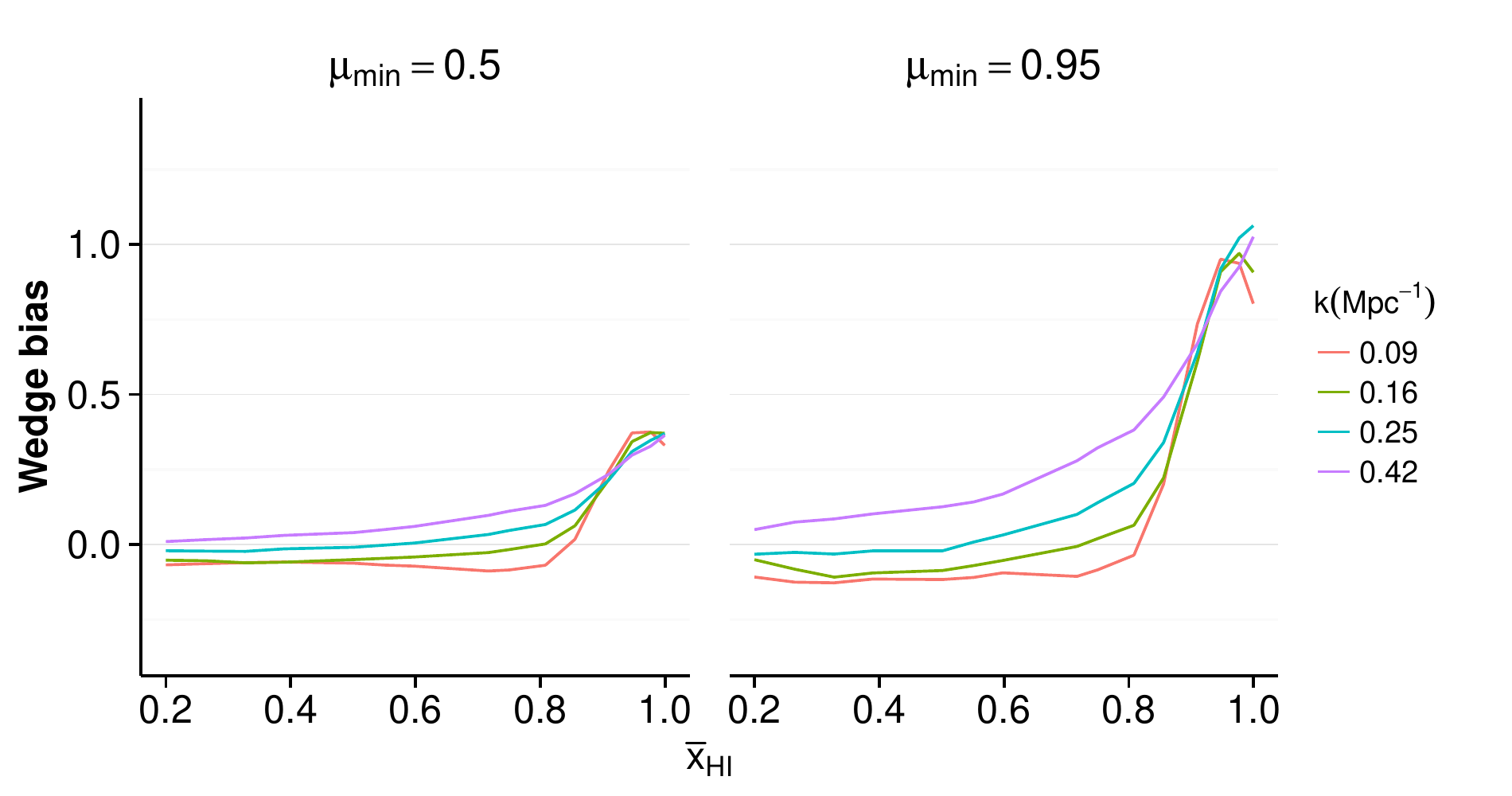}
\caption{Same as Figure \ref{fig:frac_error_fidu}, but for the massive sources model.}
\label{fig:frac_error_pl2p0}
\end{figure}
Figures \ref{fig:frac_error_fidu} and \ref{fig:frac_error_pl2p0} show the bias, i.e.\ $[\Deltawindow-\Deltasav]/\Deltasav$, when estimating the redshift-space power spectrum in the EoR window, for our two different reionization models. The bias is positive early on, with the EoR window power spectrum over-estimating the redshift-space power spectrum by around 100 per cent. At $\bar{x}_{\mathrm{HI}} \approx 0.8$, the bias turns negative and then becomes fairly insignificant in the later stages of reionization (less than 20 per cent for all $k$ modes even for $\mu_{\mathrm{min}}=0.95$). The change in sign is due to the increased anti-correlation between matter density and ionization fraction, which drives the change in the power spectrum anisotropy (see e.g. \citealt{mao12,jensen13,majumdar13,majumdar2015}).
%
%
%
%

The first, positive, peak in the bias is slightly more pronounced and occurs a little later in the fiducial model than in the massive sources model. The second, negative peak is almost absent in the massive sources model, especially on large scales. Overall, however, the difference between the two models is not dramatic, despite the massive sources model being rather extreme.
%
%

\section{Summary and discussion}
When interpreting observations of the EoR 21-cm signal, one must always be wary of the difference between the fundamental real-space signal and the observable redshift-space signal. However, measurements using foreground avoidance, which sample only some $\mu$ values, also suffer from an additional bias since the power spectrum is anisotropic in redshift space. We have quantified this bias in two different reionization simulations for a range of redshifts and spatial scales. We find that the bias is significant for higher redshifts and amplifies the difference between the real-space and redshift-space power spectra. Depending on how large a window one can obtain, the redshift-space spherically-averaged power spectrum may be overestimated by around 100 per cent. In the later stages of reionization ($\bar{x}_{\mathrm{HI}} \lesssim 0.8$), the bias instead results in an underestimation of the power spectrum. However, here the magnitude of the bias is smaller---less than 20 per cent on all $k$ modes, even for $\mu_{\mathrm{min}}=0.95$.

How can the wedge bias be accounted for in real-world observations? When performing simulations to compare to observations, the situation is rather straightforward. As long as mock observations from the simulations are produced with redshift space distortions included and using the same $\mu$ cut as the observations, the wedge bias will be accounted for automatically. A more complicated situation is when comparing observations in the EoR window to older simulation results from the literature. In this case, the wedge bias needs to be modeled. Our results indicate that the bias is not extremely sensitive to the reionization model, so it should be possible to model it to decent accuracy. In principle, it may be possible to determine directly if observed data falls in the low-bias regime by studying whether $\Deltawindow$ depends strongly on $\mu$ or not. However, this likely requires a large EoR window and low detector noise to be feasible (see \citealt{pober2015} for a related discussion).

Finally, we note that this study only presents a first, inexhaustive treatment of the wedge bias. We only show the effect for two different reionization models. The bias depends on the nature of the redshift-space distortions, which in turn depend on the reionization topology, and so the situation may be different for other reionization models. Furthermore, we have used the simplifying assumption that the intergalactic medium is heated at high redshifts, so that the spin temperature is much higher than the CMB temperature. This assumption may not hold at the highest redshifts, in which case the redshift-space power spectrum will look different \citep{ghara15}. Since the wedge bias is highest at high redshifts, it may be significantly affected by late heating. 

All the analysis in this paper was performed on data volumes with fixed redshift (``coeval volumes''). If the evolution of the signal---the lightcone effect---is taken into account, any measurement of the power spectrum will be done over a range of redshifts. The lightcone effect has been shown not to affect the anisotropy of the power spectrum at measurable $k$ modes \citep{datta14,ghara15b}, and the wedge bias will be the same as for coeval volumes.
%
%
%
%

\section*{Acknowledgements}

The research described in this paper was supported by a grant
from the Lennart and Alva Dahlmark Fund. GM is supported by Swedish
Research Council project grant 2012-4144.
AL acknowledges support from the NSF through grant AST-1109156.
ITI and KLD were supported by the Science and Technology Facilities Council [grant number STI/L000652/1]. The $N$-body simulations
were performed on the Curie system at TGCC under PRACE projects
2012061089 and 2014102339. 


\bibliographystyle{mnras}
\bibliography{refs}


\bsp	
\label{lastpage}
\end{document}